\documentclass[ras]{agutex2015}

\usepackage[T1]{fontenc} 
\usepackage{graphicx}
\usepackage{amsmath}
\usepackage{siunitx}
\usepackage{afterpage}
\usepackage{rotating}
\graphicspath {{figures/}}

\authorrunninghead{MEVIUS ET AL.}

\titlerunninghead{IONOSPHERIC STRUCTURES USING LOFAR}

\authoraddr{Corresponding author: M. Mevius, Astron, PO Box 2,
7990 AA, Dwingeloo. The Netherlands. E-mail:mevius@astron.nl}

\begin{document}

%
%

\title{Probing Ionospheric Structures using the LOFAR radio telescope}
%
%

%
%



\authors{M. Mevius\altaffilmark{1}, S. van der
  Tol\altaffilmark{1}, V. N. Pandey\altaffilmark{1},
  H. K. Vedantham\altaffilmark{1,2}, M.A.  Brentjens\altaffilmark{1}, A.G.
  de Bruyn\altaffilmark{1,2},  F. B. Abdalla\altaffilmark{3,4},
  K. M. B. Asad\altaffilmark{2}, J. D. Bregman\altaffilmark{1}, W. N. Brouw\altaffilmark{1,2},
  S. Bus\altaffilmark{2}, E. Chapman\altaffilmark{3}, B. Ciardi\altaffilmark{5},
  E. R. Fernandez\altaffilmark{2}, A. Ghosh\altaffilmark{2}, G. Harker\altaffilmark{3},
  I. T. Iliev\altaffilmark{6}, V. Jeli\'{c}\altaffilmark{1,2,7},
  S. Kazemi\altaffilmark{8},
  L. V. E. Koopmans\altaffilmark{2}, 
  J. E. Noordam\altaffilmark{1},
  A. R. Offringa\altaffilmark{1}, A. H. Patil\altaffilmark{2}, R.J. van
  Weeren\altaffilmark{9}, S. Wijnholds\altaffilmark{1}, S. Yatawatta\altaffilmark{1}, S. Zaroubi\altaffilmark{2}}
\altaffiltext{1}{Astron, PO Box 2, 7990 AA, Dwingeloo, The Netherlands.}
\altaffiltext{2}{Kapteyn Astronomical Institute,University of Groningen,PO Box
  800, 9700 AV, Groningen, The Netherlands.}
\altaffiltext{3}{Department of Physics \& Astronomy, University College
  London, Gower Street, London WC1E 6BT, UK.}
\altaffiltext{4}{SKA SA, 3rd Floor, The Park, Park Road, Pinelands, 7405, South Africa.}
\altaffiltext{5}{Max-Planck Institute for Astrophysics, Karl-Schwarzschild-Strasse 1, D-85748 Garching bei M\"unchen, Germany.}
\altaffiltext{6}{Astronomy Centre, Department of Physics \& Astronomy, Pevensey II Building, University of Sussex, Brighton BN1 9QH, UK.}
\altaffiltext{7}{Ru{\dj}er Bo\v{s}kovi\'{c} Institute, Bijeni\v{c}ka cesta 54,
  10000 Zagreb, Croatia.}
\altaffiltext{8}{ASTRON~\&~IBM Center for Exascale technology, Oude Hoogeveensedijk 4, 7991 PD Dwingeloo, the Netherlands.}
\altaffiltext{9}{Harvard-Smithsonian Center for Astrophysics, 60 Garden Street, MA 02138, USA.}






%
%


\keypoints{\item LOFAR is able to measure differential TEC values with an
  accuracy better than 1 mTECU.
\item The diffractive scale is an easily obtained single number indicating the ionospheric quality of a radio interferometric observation.
\item The ionospheric phase structure functions of most nights show a
  spatial anisotropy that is in many cases Earth magnetic field aligned.
}


%
%


\begin{abstract}

LOFAR is the LOw Frequency Radio interferometer ARray located at mid-latitude
($52^{\circ} 53'N$). Here, we present
results on ionospheric structures derived from 29 LOFAR nighttime observations
during the winters of 2012/2013 and 2013/2014.  We show
that LOFAR is able to determine differential
ionospheric TEC values with an accuracy better than 1 mTECU over distances
ranging between 1 and 100 km. For all observations the power law behavior of the
phase structure function is confirmed over a long range of baseline lengths, between $1$ and
$80$ km, with a slope that is in general larger than the $5/3$ expected
for pure Kolmogorov turbulence. The measured average slope is $1.89$ with a
one standard deviation spread of $0.1$. The diffractive scale, i.e. the length scale where the
phase variance is $1\, \mathrm{rad^2}$, is shown to be an easily
obtained single number that represents the ionospheric \emph{quality} of a
radio interferometric observation. A small diffractive scale is equivalent to
high phase variability over the field of view as well as a short time
coherence of the signal, which limits calibration and imaging quality. For the studied observations
the diffractive scales at  $150$ MHz vary between $3.5$ and $30\,$ km. 
A diffractive scale above $5$ km, pertinent to about $90 \%$ of the
observations, is considered sufficient for the high dynamic range imaging
needed for the LOFAR Epoch of Reionization project.
For most nights the ionospheric irregularities were anisotropic, with the structures being aligned with the Earth magnetic field in about
$60\%$ of the observations.

\end{abstract}

%
%

%

\begin{article}

%
%

\section{Introduction}
With the arrival of long baseline radio interferometric arrays such as LOFAR
\citep{haarlem}, low
frequency radio astronomy has reached a new era. In radio astronomy low
frequency refers to frequencies between a few tens of MHz and a few hundred MHz.
LOFAR operates between 30 and 250 MHz. One of the scientific cases for LOFAR is the
measurement of the redshifted 21-cm emission line of neutral hydrogen from the
epoch of reionization (EoR) \citep{haarlem}. This
EoR signal is expected to lie many orders
of magnitude below the foreground astrophysical emission in a single
observation. Thus, high dynamic range imaging is needed to extract the EoR
signal. The required precision
poses new challenges in calibration and imaging of the data. Ionospheric
propagation delays are a major contributor to phase errors at low radio
frequencies. Residual effects in the data due to ionospheric phase errors on
the bright astrophysical foreground emission can pose a significant challenge to
EoR experiments. It is for these reasons that we started a program to investigate the ionospheric
disturbance in the LOFAR EoR observations. While the main goal of such a program is to
remove the effects of the ionosphere, investigating the ionospheric phases also
reveals a wealth of information on physical processes in the ionosphere, which
is interesting in its own right. In addition, knowledge of the
characteristics of the ionosphere gives improved estimates of the residual
ionospheric \emph{speckle noise} in the radio interferometric images after
calibration \citep{koopmans2010,vedantham}. In this paper we focus on the
ionospheric information extracted from LOFAR EoR calibration data that is
relevant for assigning an ionospheric \emph{quality} to a given
radio interferometric observation.

An electromagnetic signal with frequency $\nu$ passing through the ionosphere
undergoes an additional phase
shift that is to first order equal to:
\begin{equation}\label{eq:phasedelay}
\Delta\phi=-8.45\left(\frac{TEC}{1{\rm\,TECU}}\right) \,\,
\left(\frac{\nu}{1{\rm\,GHz}}\right)^{-1} \,{\rm rad}
\end{equation}
where TEC is the integrated electron density along the line of
sight in TEC units and $1\,\mathrm{TECU}=10^{16} m^{-2}$.  The above equation is only
valid for
observing frequencies well above the maximum plasma frequency ($\approx 10\,$ MHz). From the spectral dependence in equation
\eqref{eq:phasedelay} it is clear why the ionosphere is a major source of
calibration errors especially at low
frequencies. 
An interferometer only measures phase differences, therefore, to first order, its signal is
only distorted if the ionospheric electron column densities above the two elements of an
interferometer differ. Early experiments have shown that the
ionosphere can be considered as a turbulent
medium with scales over a long range of distances
\citep{wheelon}. Consequently, the variance of the differential TEC is
expected to follow a power law with
the distance between two points in the ionosphere. It is the purpose of this paper to study
this power law behavior and its slope over the long range of distances defined by the available
baseline lengths of the Dutch LOFAR stations. 

Ionospheric phases can be partly removed from the data using 
self calibration \citep{selfcal}. In self calibration, a model of the sky and
the instrument is used to predict the radio interferometric
visibilities. Calibration parameters, including both amplitude and phase
corrections for each element, are then determined by fitting the
predicted visibilities to the data. The sky model is updated in an iterative
sequence. Traditional self calibration uses a single set of parameters
for all directions and does not take into account  the angular variation in
ionospheric distortions over the field of view. Direction-dependent
calibration \citep{smirnov}
can take care of this, but it needs a good model of the radio sky and it can
only be done for a limited number of directions. Methods to interpolate
between the direction dependent ionospheric parameters and subsequently apply them in other
directions, are for example field based calibration \citep{cotton} or Source Peeling and Atmospheric Modeling (SPAM) \citep{intema}. In self calibration the sky
model is extracted from the data themselves. For a high resolution sky model, the
longest baselines are needed, which suffer most from
ionospheric distortions. In effect, residual ionospheric  phase noise is inevitable,
and good knowledge of the structure of the ionosphere helps in understanding
this source of stochastic errors.

Most of our knowledge of the large scale ionosphere is derived from measurements from dedicated
experiments with ionosondes and GPS satellites and receivers.  Measurements of
ionospheric structure has been done by radio telescopes before, for example
traveling ionospheric disturbances (TIDs) \citep{velthoven,spoelstra,helmboldt}, turbulent
like fluctuations \citep{spoelstra,cohen} and plasma-spheric irregularities \citep{jacobson,helmboldt,mwa}. LOFAR can contribute to these measurements in a unique way. The
wide bandwidth of LOFAR observations in both its low band (30-80 MHz) and high band (110-190
MHz) observing mode, allows good separation  of the ionospheric effects from other errors that
have a different frequency dependence. Also, the LOFAR layout facilitates probing
ionospheric structures on a large range of scales. The dense inner core gives
an instantaneous imprint of the small scale structures  ($\sim 2$ km),
whereas with the remote stations, LOFAR is sensitive to ionospheric structures up
to 100 km in size. LOFAR also has stations in Germany, the UK,
  France, Sweden and recently added Poland, building up baselines of more than 1000 km.  Apart from exploiting LOFAR's array layout, its wide field of view and the
simultaneous multi-beaming capability facilitates studies of ionospheric
structures over a large range
of spatial scales in a complementary way. In this case, the
phase distortions are not measured towards a single source, but towards a large
number of sources distributed over the large field of view (FOV) of one or many simultaneous
beams. 

In this paper we will exploit
LOFAR's phase solutions in the direction of a single bright and dominant calibrator to probe
ionospheric structures on spatial scales corresponding to LOFAR's baselines,
where we restrict ourselves to data from \emph{Dutch} LOFAR. We thus use the
results of traditional self calibration in a single direction. The many-source approach will be the topic of a subsequent paper. We will globally discuss the effect of
ionosphere on the quality of radio interferometric images. The diffractive
scale is introduced as a single number representation of the ionospheric
quality of an observation. A more quantitative approach using the diffractive
scale to decide on calibration strategies, will be left for future work. 

In the second section we will discuss our dataset and in the third we outline how we collected ionospheric information
from self calibration solutions. Section 4 deals with the framework of
ionospheric spatial variability. The two dimensional structure function will be
introduced here.  In the fifth section the result of fitting this ionospheric
structure function on data of many nights is shown and the
ionospheric phase structure function will be correlated with image noise. Here
we will also discuss anisotropy and elongation of structures along the magnetic field
lines, which has been observed for many nights. Discussion of the
results will follow in the final conclusion.

\section{Data Description}\label{sec:data}

\begin{figure}
\noindent\includegraphics[width=\hsize]{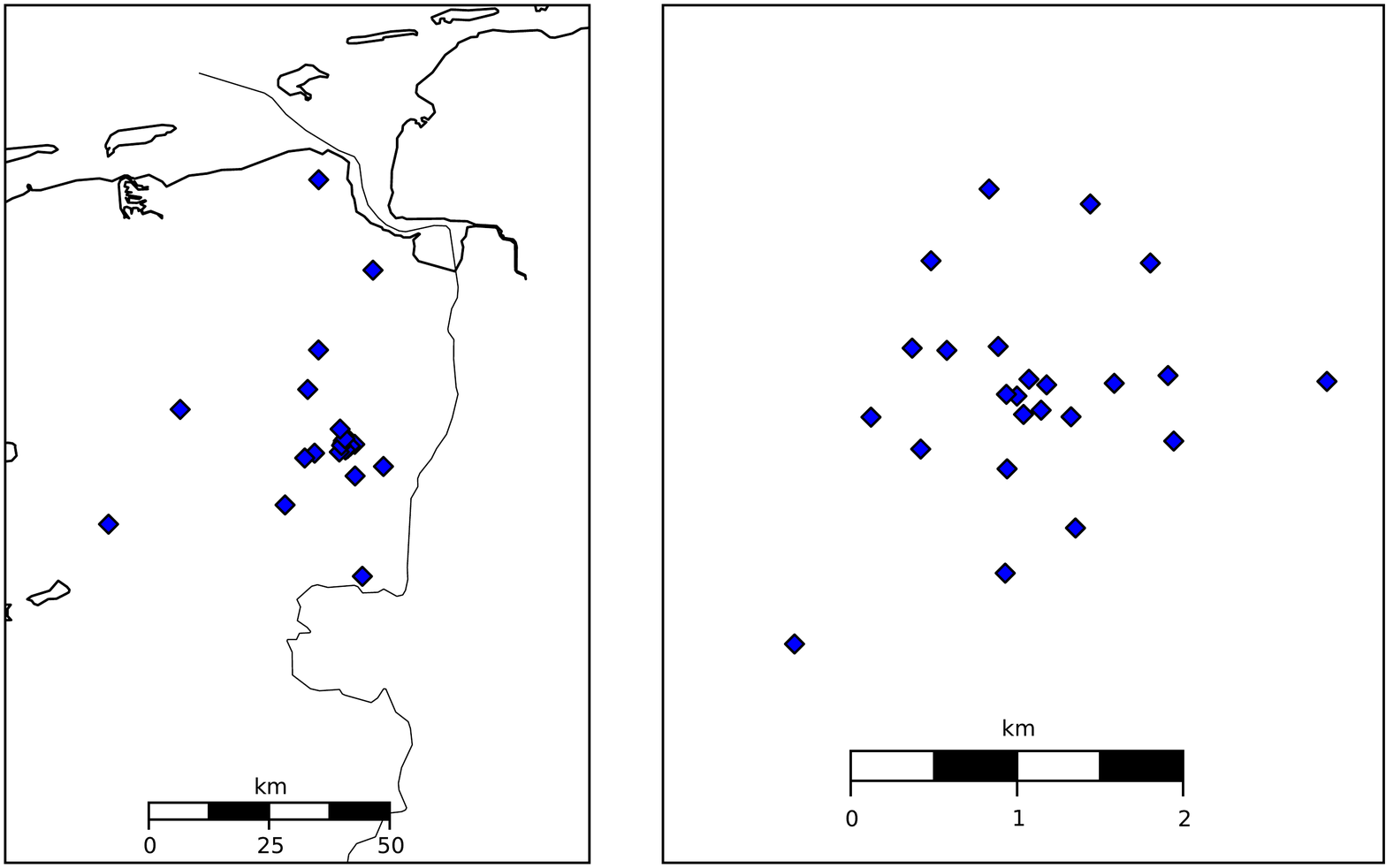}
\caption{Layout of the LOFAR Stations. Left: all Dutch stations. Right: The
  central core of LOFAR.}
\label{fig:lofar_layout}
\end{figure}

The center of LOFAR is located in the Netherlands at mid latitude ($52^{\circ}
53' N,6^{\circ} 52' E$).
The analysis in this paper is
performed using data between 110 and 190 MHz from LOFAR's High Band Antennas
(HBA). It uses all baselines formed from LOFAR's core and remote stations,
giving projected baselines ranging from 30 m to 100 km. The LOFAR core
HBA stations are arranged in two fields each with a diameter of 31 m. For the EOR observations these are treated as individual stations at the correlator, resulting in a
total of 48 core stations, of which 46 were used in the analysis. The number
of remote stations at the time of observations was 13 for most, the 14th station
was added during 2013. The Full HalfWidth Maximum (FWHM) of the stations varies between
4.8 and 3.2 degrees in the given frequency range.  Figure
\ref{fig:lofar_layout} shows the layouts of all Dutch LOFAR stations and the
LOFAR core.

For our analysis we used many nights with data collected for the Epoch
of Reionization
 project. In the LOFAR-EoR project the data of several
hundreds of hours of observations of the same field are combined to extract
the signal. The two main fields that are being observed are one in the direction of the
North Celestial Pole and the other with the phase center at the bright quasar
3C196 (J2000:RA 08h13m36s,Dec +48d13m03s). We chose the latter field for our analysis, since the
bright calibrator in the center of the field eases the extraction of the
ionospheric information. The EoR project requires nighttime
observations. Therefore, in the current analysis, we only study nighttime characteristics of the ionosphere. Also, since 3C196 is only
above the horizon during nighttime in the winter, our observations are
restricted to these months. 

We selected the first 29 observations of the 3C196 field. Of these, 26 were
recorded in the winter of 2012/2013, the last three are from 2013/2014. Most
of the observations last for 8 hours, centered on the meridian transit of 3C196. There are also six 6 hour observations. For comparison and to avoid
systematic effects that are mainly present at low elevations, all results
presented here use only the middle 6 hours of the observations, unless specified otherwise. The exact epochs and durations of the
observations are listed in table \ref{tab:fitresults}. 

The data consist of 380 subbands, each with a width of 195.3 kHz, covering the frequencies between 115 and 189
MHz. The raw data in each subband were recorded with an integration time of 2s
and 64 channels per subband (each with a width of about 3 kHz), which after flagging
for Radio-Frequency Interference (RFI) using the AOFlagger algorithm \citep{offringa2010,offringa2012}
were averaged to the final data product with a resolution of 10 seconds  and 1 channel per
subband. The first and last two channels of the original data were
discarded, resulting in a width of 183.1 kHz per subband. After flagging and averaging, the data were calibrated with the LOFAR
Blackboard Self calibration (BBS) system \citep{bbs}. During calibration, the complex gains
are fitted by minimizing the difference between the recorded visibilities and
the model visibilities. The model visibilities are generated using a model of the sky and the antenna pattern. The resulting calibration gains contain the
effect of any distortions between the emitter and the antenna, among which are
the residual unmodeled antenna gain pattern, instrumental errors and atmospheric and ionospheric
effects \citep{smirnov}. Our sky model consists solely of 4 components of the
dominant source
3C196 [\emph{Pandey V.N. private communication}]. We ignored the weaker sources in the field, since 3C196
is about 14 times brighter than the second brightest source. The
effects of the \emph{missing} sky in our model
can be seen as a second order effect in the phase solutions, which will be
discussed in more detail in section \ref{section:phasestructure}. 
The effect of the primary beam was taken into account using the standard LOFAR
beam model \citep{bbs}.


\section{Obtaining  ionospheric information from calibration phases}\label{section:clocktec}
The main contribution of ionospheric propagation at LOFAR
frequencies is a dispersive delay, showing up in the calibration phases. However,
the interferometric calibration phases also contain other effects. In
particular, since the remote stations are not on a
common clock as the core, a main source of (time-varying) phase errors are the drifting
clock errors. The clock errors can be as large as 200 ns, with a drift
rate in the order of $1e^{-12}$.  Other, smaller, effects include cable
reflections, beam model errors, tropospheric delay fluctuations and the imperfect sky model. 

It is possible to distinguish clock from ionospheric phases by using the wide
frequency range and the difference in frequency behavior of the two
effects. The phase error between station $i$ and $j$ is:
\begin{equation}\label{eq:clocktec}
\delta\phi_{ij}(\nu)= (2\pi\cdot\delta\tau_{ij}\cdot\nu  - C_1\cdot\delta
TEC_{ij}/\nu)\,\,\,\,\mathrm{rad},
\end{equation}
 where $C_1 \approx 8.45e^9 \mathrm{m^{2}s^{-1}}$ is the ionospheric
 conversion, $\nu$  the signal frequency in Hz and $\delta TEC_{ij}$ the
difference in integrated ionospheric electron content in TECU between the line
of sights to stations $i$ and $j$. The relative timing error on the clocks at the
stations $i,j$ is $\delta\tau_{ij}$ (s). In order to extract the
TEC information from the calibration phases, we selected the phase solutions
of 31 subbands uniformly distributed over a 115-175 MHz range and fitted
function \eqref{eq:clocktec} to the data. We limited the number of subbands used
in our analysis to about $8\%$ of the available data for reasons of computational cost. Because the final accuracy of the TEC values is limited
by systematic errors, as will be shown in section
\ref{section:phasestructure}, this has negligible influence on the accuracy of our results.

Since $\delta\phi_{ij}$ can only be measured up to a $n\cdot2\pi$ ambiguity, the phase data were first
unwrapped. Unwrapping can be done if the parameters $\delta\tau_{ij}$ and
$\delta TEC_{ij}$  are known to a reasonable precision a priori. Best initial estimates of $\delta\tau_{ij}$
and $\delta TEC_{ij}$ were found for the first time slot of an observation by
searching over a large range of possible solutions to find the
best fitting match. For subsequent time slots the parameters were initialized
with the solution of the previous time slot and the phases unwrapped
accordingly before performing the fit. This is possible since  we do not
expect the station clocks and/or the differential TEC to vary substantially within a single time slot of 10
seconds. The accurate phases due to the high signal to noise ratio of the
calibrator ensures that the solutions stay in the same local minimum ($2\pi$
interval) over time. A full $2\pi$ phase wrap corresponds in the observed
frequency range to a jump in the TEC and clock
value of $\sim\,0.05\,$TECU and $\sim\,3$ ns, respectively. No such jumps were found when checking the time variation of
the fitted parameters. The maximum absolute difference between two time slots for all
observations is $0.029\,$ TECU and $0.9\,$ ns for the TEC and clock solutions, where on
average it is $0.0015$ TECU and $0.05$ ns for the longest
baselines.

\begin{figure}
\noindent\includegraphics[width=\hsize]{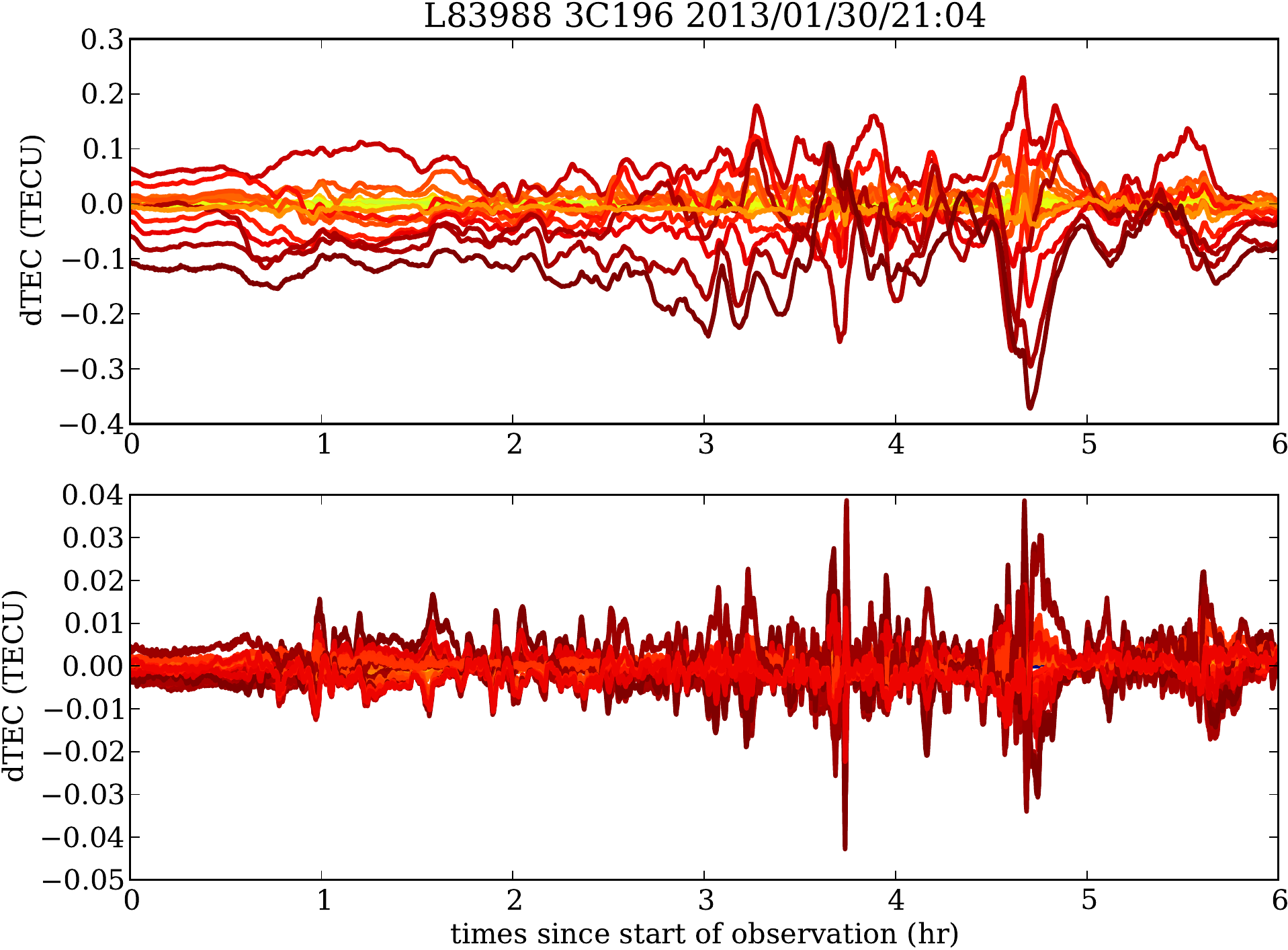}
\caption{Differential TEC of LOFAR stations with respect to the central
  station versus time, single observation. The color
  coding is corresponds to the baseline length, with longer
  baselines having darker colors. top: All stations. bottom: Only core
  stations (maximum baseline 2 km).}
\label{fig:TECvstime}
\end{figure}

After an initial iteration of the clock/TEC separation fit,
the time averaged spatial correlation of differential TEC values was checked per observation by fitting a linear 2 dimensional
polynomial over the time averaged TEC values projected on the positions of the
different stations. Remaining $2\pi$ phasewraps (by construction constant over
the full observation) could be detected this way as well as a small constant (both
in frequency and time) phase offset per station. Such a phase offset is
typically introduced by the station calibration solutions, that are applied
on-the-fly to the data prior to interferometric correlation. After these
corrections, final parameter values were determined in a second iteration of the clock/TEC
separation. A typical example of the fitted differential TEC values is shown in figure
\ref{fig:TECvstime}  both for the full array and for the core
stations. The TEC values are shown with the
central station CS001HBA0 as a reference. The color scale corresponds to
baseline length. Temporal
ionospheric variations that are spatially correlated from
station to station are clearly observed even on the short ($\sim\,1\,$km) baselines.

The accuracy of the TEC solutions that can be reached with this method is
limited by ignoring second order phase effects. Especially effects that are not
linear in frequency, and thus cannot be absorbed in the clock solutions, will
be partially absorbed in the TEC solutions. Examples of such phase errors are
cable reflections, sky and beam model errors. 

\section{Ionospheric Structure Function}\label{section:phasestructure}

The spatial structure of a medium like the ionosphere can be characterized by its power
spectrum, or equivalently, its Fourier inverse -- the phase correlation
function. In practice, it is convenient to measure the spatial correlation
in terms of the phase structure function \citep{tol2009}, defined as:
\begin{equation}\label{eq:phasestructure}
D(r)= \left<\left(\phi(r')-\phi(r'+r)\right)^2\right>
\end{equation}
For Kolmogorov turbulence, the phase structure function takes the form of a
power-law in the inertial range of turbulence:
\begin{equation}\label{eq:Kolmogorov}
D(r) = \left(\frac{r}{r_{\rm diff}}\right)^\beta,
\end{equation}
where $r_{\rm diff}$ is the spatial scale over which the phase variance is
$1$ rad$^2$, and is referred to as the \emph{diffractive scale} \citep{narayan1992}. The index $\beta$ is
equal to $5/3$ for pure Kolmogorov turbulence. 

In order to measure the structure function using the LOFAR data, we determine
the variance per baseline of the differential TEC as shown in figure
\ref{fig:TECvstime}. From figure \ref{fig:TECvstime} it is clear that the time
average of the differential TEC for two stations is not zero. In fact there is a large
North South TEC gradient over the array that is visible in all observations
and is merely a result of the very large scale global ionospheric structure. By calculating
the phase variance, and thus using the mean subtracted values, we implicitly filter out
this global structure. We converted the
slant TEC (sTEC) values to vertical TEC (vTEC) by dividing with the slant factor, assuming a single
layer ionosphere at 300 km, taking into account the zenith angle in the
direction of the source. This correction factor is calculated as:
\begin{equation}\nonumber
vTEC=cos(\alpha^\prime)\cdot sTEC,
\end{equation}
\begin{equation}\label{eq:airmass}
\alpha^\prime=asin\left(\frac{R_{Earth}}{R_{Earth}+h}\cdot sin(\alpha)\right).
\end{equation}
With $R_{Earth}$ the Earth radius and $h$ the altitude of the single layer
approximation of the ionosphere and $\alpha$ the zenith angle. The dependence
of this correction on the exact altitude of the single layer is minor. 

\begin{figure}
\noindent\includegraphics[width=\hsize]{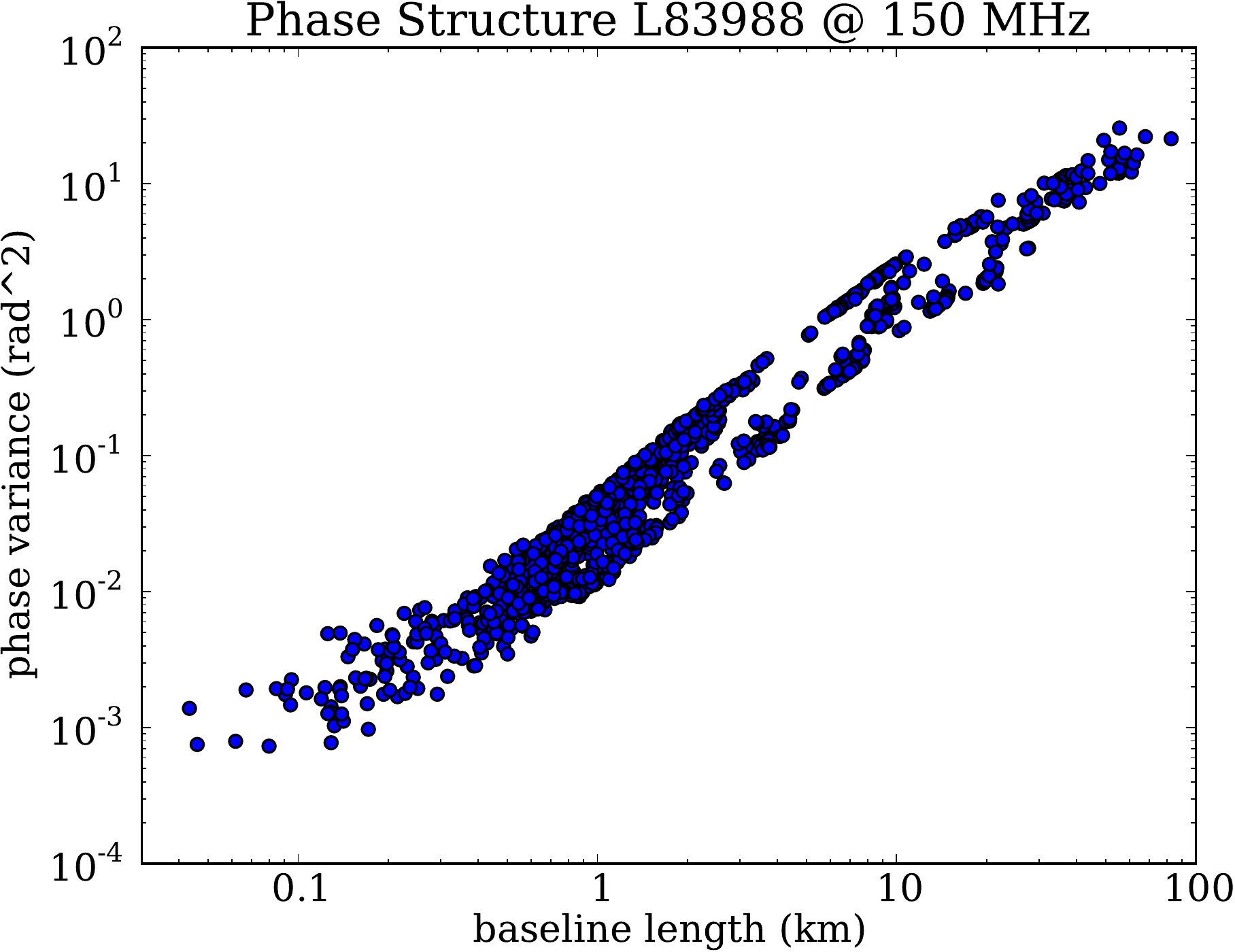}
\caption{Phase structure function of the same observation as in figure
  \ref{fig:TECvstime}. The differential TEC values are converted to phases at
  150 MHz.}
\label{fig:unbinned_structure}
\end{figure}

The vertical differential TEC values
were converted to a differential phase at 150 MHz for comparison with other experiments using
equation \eqref{eq:phasedelay}. We use
the time average of the phase variations to estimate the ensemble
average. The observations are 6 hours long, a timescale much longer than the coherence scale. In these 6  hours we are tracking a source,
corresponding to tracking the ionosphere over a projected distance of $\sim 300$ km on an ionospheric
layer at $300$ km height. At the same time ionospheric structures are
moving. TIDS, for example, are moving with a typical speed of a few hundred
km/hr. In general this increases the size of the sampled space, although one should keep
in mind that the propagation direction of the ionospheric structures could coincide
with that of the source. For the reasons mentioned above, we sample enough independent data points
on the ionospheric screen to
allow the ergodic theorem to hold.  An example of the un-binned
spatial phase structure function of a typical observation, the same as used for
figure \ref{fig:TECvstime}, is shown in figure
\ref{fig:unbinned_structure}. 

The figure shows some typical features. First, for a
large range of baseline lengths, between 1 and 80 km, the power law behavior is apparent. There is a
hint of a turnover at the very long baselines $(\>\sim 80$ km), which may
represent the outer scale of turbulence at which the structure function is
expected to saturate. However, it could also likely be caused by regular
structures, for example a traveling wave with a wavelength of about twice that of the length at the
turnover point ($\sim 150$km). The contribution of phase tilt of such a wave to the phase
structure function would be a power law with power 2.0\citep{wandzura1980},
larger than the $5/3$ for Kolmogorov turbulence. Inclusion of LOFAR's (longer)
international baselines in the analysis will yield an unambiguous measurement
of the outer scale, which we have not pursued here. 
\begin{figure}
\noindent\includegraphics[width=\hsize]{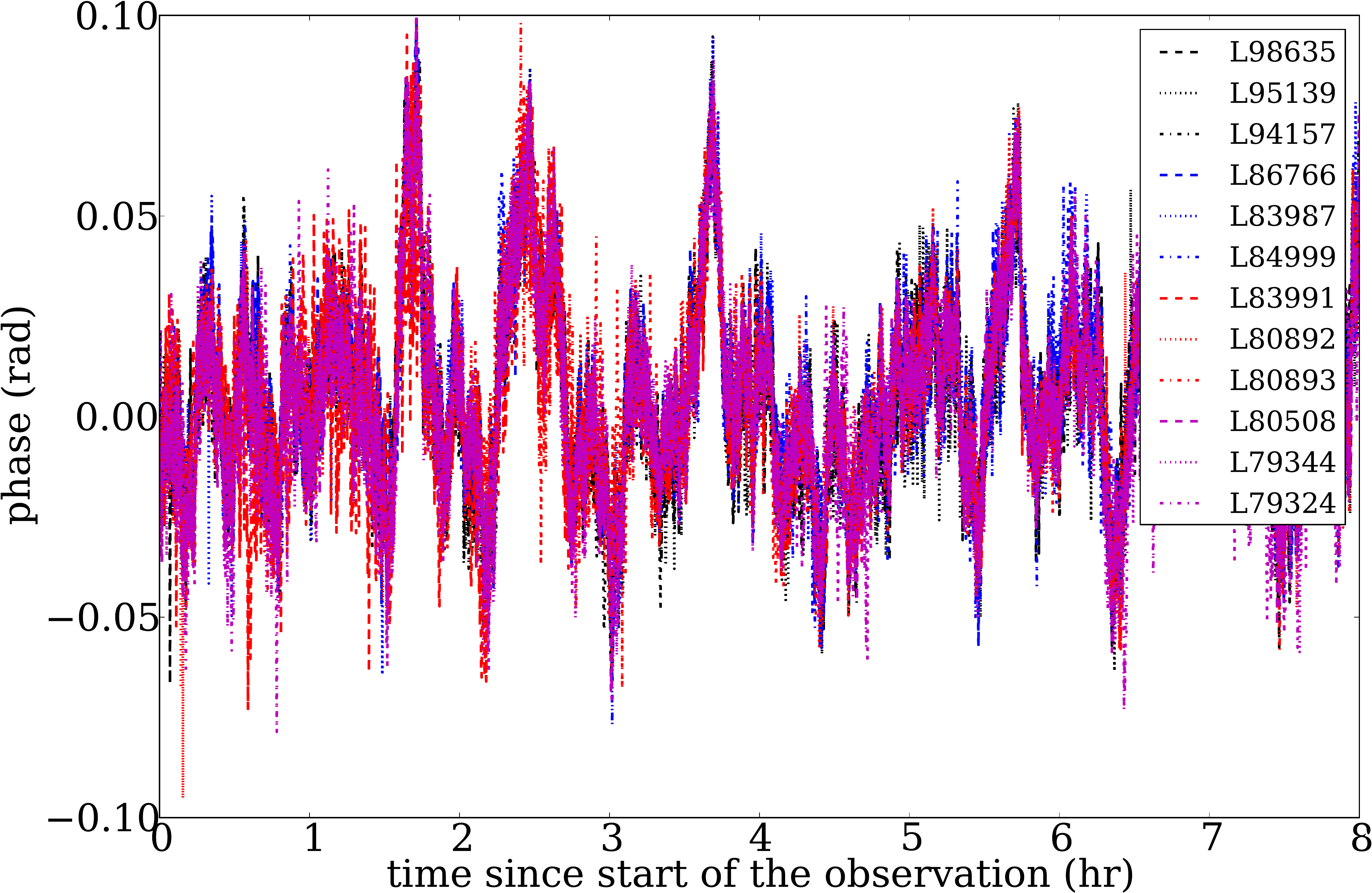}
\caption{Phase residuals versus time after clock/TEC separation, single
  subband. The different lines correspond to different observations. There is
  a striking correlation between the residuals of different observations,
  suggesting a common cause of systematic errors.}
\label{fig:residuals}
\end{figure}
 
\begin{figure}
\noindent\includegraphics[width=0.8\hsize]{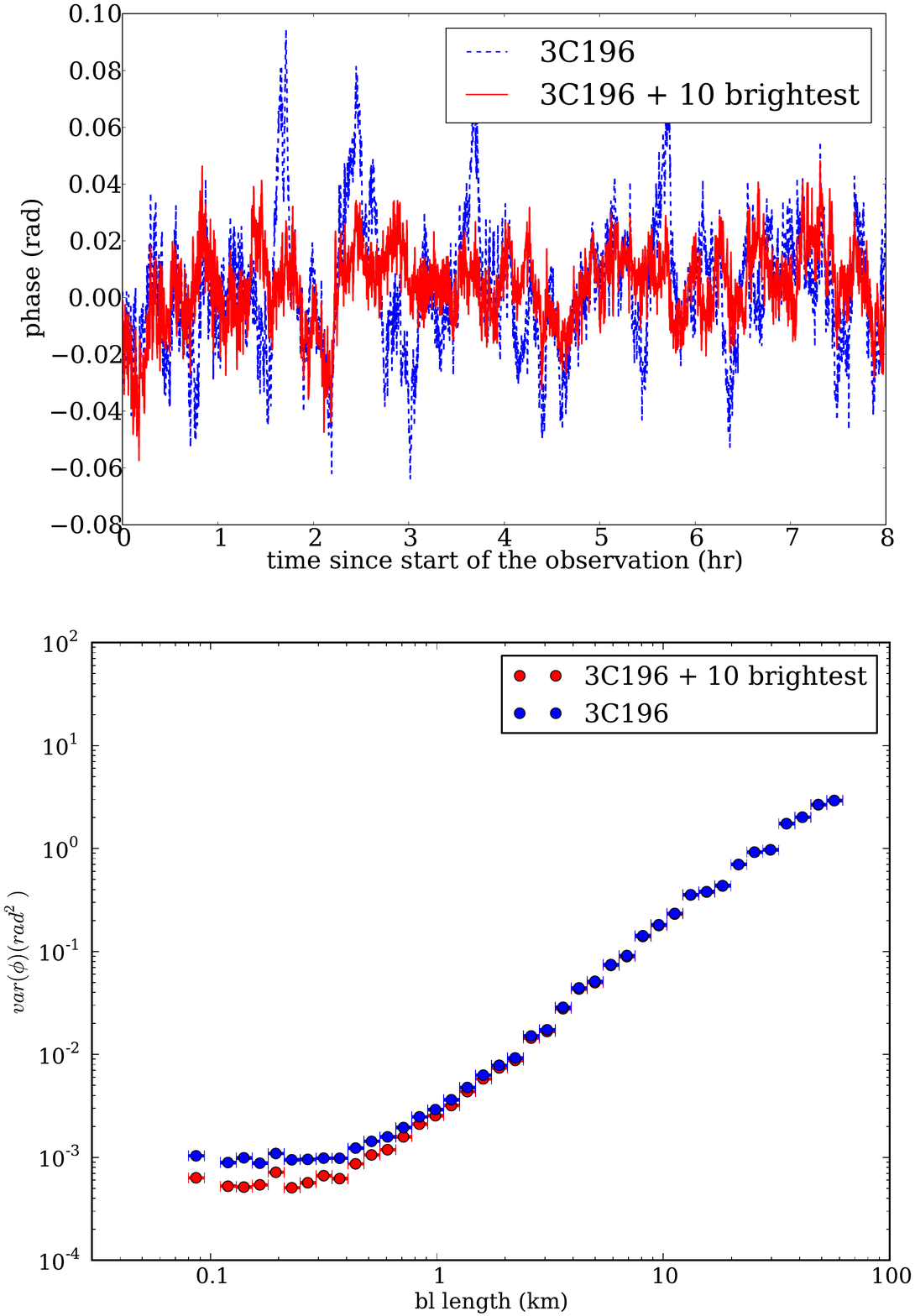}
\caption{Top: Phase residuals after clock/TEC separation. Comparison of the
  residuals using a single source model and a more complete sky model with 10
  additional sources during calibration. Bottom: Difference in phase structure
  for the two different sky models. Only the middle 4 hours of the observation
  were taken into account. The noise floor at shorter baselines is about a
  factor 2 lower for the improved model.}
\label{fig:residuals_10_sources}
\end{figure}

The turnover at the shorter baseline lengths in figure
\ref{fig:unbinned_structure} is due to the presence of a noise floor. As discussed in
section \ref{section:clocktec}, the second order contributions to the
calibration phases, not taken into account during clock/TEC
separation, lead to small systematic variations in the fitted TEC values. This
can be studied by examining the residuals of the fit of the function in equation \eqref{eq:clocktec} to the
calibration phases. The time variation of these residuals  is shown in figure \ref{fig:residuals} for a single baseline and a
single subband of a large range of observations. There is  a strong
correlation between the residuals of different observations, suggesting a
common systematic error. Since all observations are aligned
in local sidereal time, either the incompleteness  of the sky model or beam
model errors are good candidates for these residual calibration phases. See
for example \citet{wijnholds2016} for a more elaborate discussion of the effect
of missing flux on calibration errors. For
one observation we redid the calibration and clock/TEC fitting with the 10
brightest sources in the field added
to the sky model. The resulting residuals, shown in figure
\ref{fig:residuals_10_sources}, are indeed smaller
and different in structure. We also show in figure \ref{fig:residuals_10_sources}
that the noise floor of the phase structure function for this
particular observation indeed drops when including more sources in the model,
when only the middle 4 hours of the observation were used. However, if we
include the full 8 hours of the observation we noticed a systematic effect especially at the start and end of
the observation, when the elevation angles are lower, that could not be reduced
by improving the sky model.  This source of systematic noise at low elevations
could probably be attributed to an imperfect model of the primary beam. If not properly taken into account, these systematic effects in the phases will lead to small systematic errors in the Clock/TEC
separation. Assuming the turnover at the short baselines in the structure function can be completely
attributed to these systematic errors, we can estimate the
uncertainty on the TEC values from the noise floor. Noise will add an
additional term to the phase structure function. We chose to use a single
constant term for the noise, although, in principle, the systematic
noise could be spatial correlated. To check the validity of this simplification
we measured the structure function of the difference between 
the independently fitted TEC of both polarizations, which showed a flat spectrum. The expression for the estimated phase structure becomes:
\begin{equation}\label{eq:noiseKolmogorov}
D(r) = \left(\frac{r}{r_{diff}}\right)^\beta+\sigma^2,
\end{equation}
which describes the phase structure for an isotropic medium up to the turnover
point at large scales.

Another typical
feature  of the  phase structure function  is the band-like structure, of which an example in figure \ref{fig:unbinned_structure}. This structure is seen in many observations
and was found to be dependent on
the orientation of the  baseline. The fact that the phase structure has
a different scale ($r_{diff}$) for different directions, although the slope
is more or less equal, suggests that the ionospheric irregularities are anisotropic. Like the
turnover point at long baseline lengths, this
envelope structure is also consistent with a contribution from large wavelike structures  that propagate in the
direction where the smallest diffractive scales are measured. The anisotropy of the ionospheric
structure has been observed before \citep{wheelon,spencer,singleton}. It can be
taken into account in the structure function by making the function two
dimensional:
\begin{equation}\nonumber
 E(\mathbf{r})=(\mathbf{r}^\top \mathbf{R}^\top \Sigma 
 \mathbf{R}\mathbf{r})^{\beta/2}+\sigma^2
\end{equation}
\begin{equation}
\Sigma = diag\left(\left(\frac{1}{r_{diff}^{maj}}\right)^2,\left(\frac{1}{r_{diff}^{min}}\right)^2\right),
\label{eq:2DKolmogorov}
\end{equation}


with  $r_{diff}^{maj},r_{diff}^{min}$ the diffractive scales of the major and
minor axes and $\mathbf{r} = (r_x,r_y)$ the vector of baseline lengths
projected on an appropriate orthogonal frame (we chose $r_x$ to be EW and $r_y$ to be NS oriented). $\mathbf{R}$ is the
2$\times$2 rotation matrix of angle $\alpha$, the orientation of the major axis within this frame (North to East).
A qualitative discussion on the anisotropy and its orientation can be found in section \ref{sec:anisotropy}. 

We constructed the phase structure function for all observations under
consideration and fit for parameters $r_{diff}^{maj},r_{diff}^{min},\alpha,\beta$ and
$\sigma$ in equation \eqref{eq:2DKolmogorov} to determine the
characteristics of the ionosphere. As will be discussed in section
\ref{sec:image_noise}, the diffractive scale
is a good candidate to quantify
the ionospheric \emph{quality} of a night with a single number. In order to assign a
single measure for the ionospheric quality of the night, we used the simple 1D
function in equation \eqref{eq:noiseKolmogorov} to get an estimate of the average diffractive
scale $r_{diff}$, where, for stability, we fixed the value of $\beta$ to the result of
equation \eqref{eq:2DKolmogorov} to the data. 

The time averaging can lead to systematic errors in the determination of the
structure function parameters. Large peaks in the
differential TEC values, for example, can have a major impact on the variance. We investigated the systematic errors due to the time averaging in the
following way. For each observation we generated 5 independent subsets with random time
sampling. The fit was performed for all subsets and we calculated the
variance of the five
values per parameter. These variances were in general much larger than the statistical
errors from the fit, indicating that there are indeed systematic uncertainties. These standard deviations were quadratically added to the covariance errors from the fit
of the full dataset to get a conservative estimate of the total uncertainty
of the fitted parameters.

\begin{figure*}[!htbp]
\noindent\includegraphics[width=\textwidth]{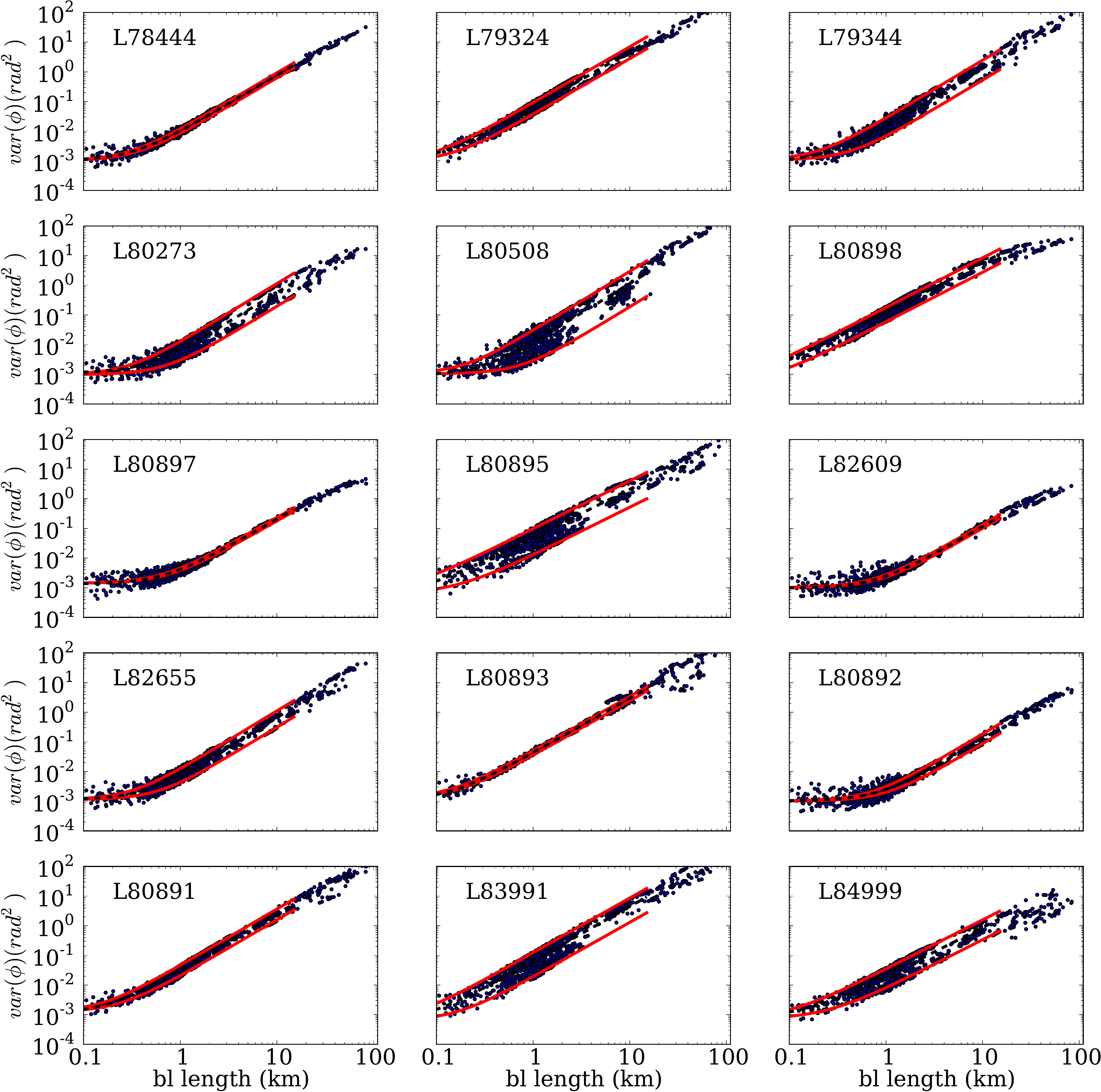}
\caption{Structure functions of all observations. The lines are the major and minor
  axis projections of the fitted function \eqref{eq:2DKolmogorov}.}
\label{fig:structure_all1}
\end{figure*}

\begin{figure*}[!htbp]
\noindent\includegraphics[width=\textwidth]{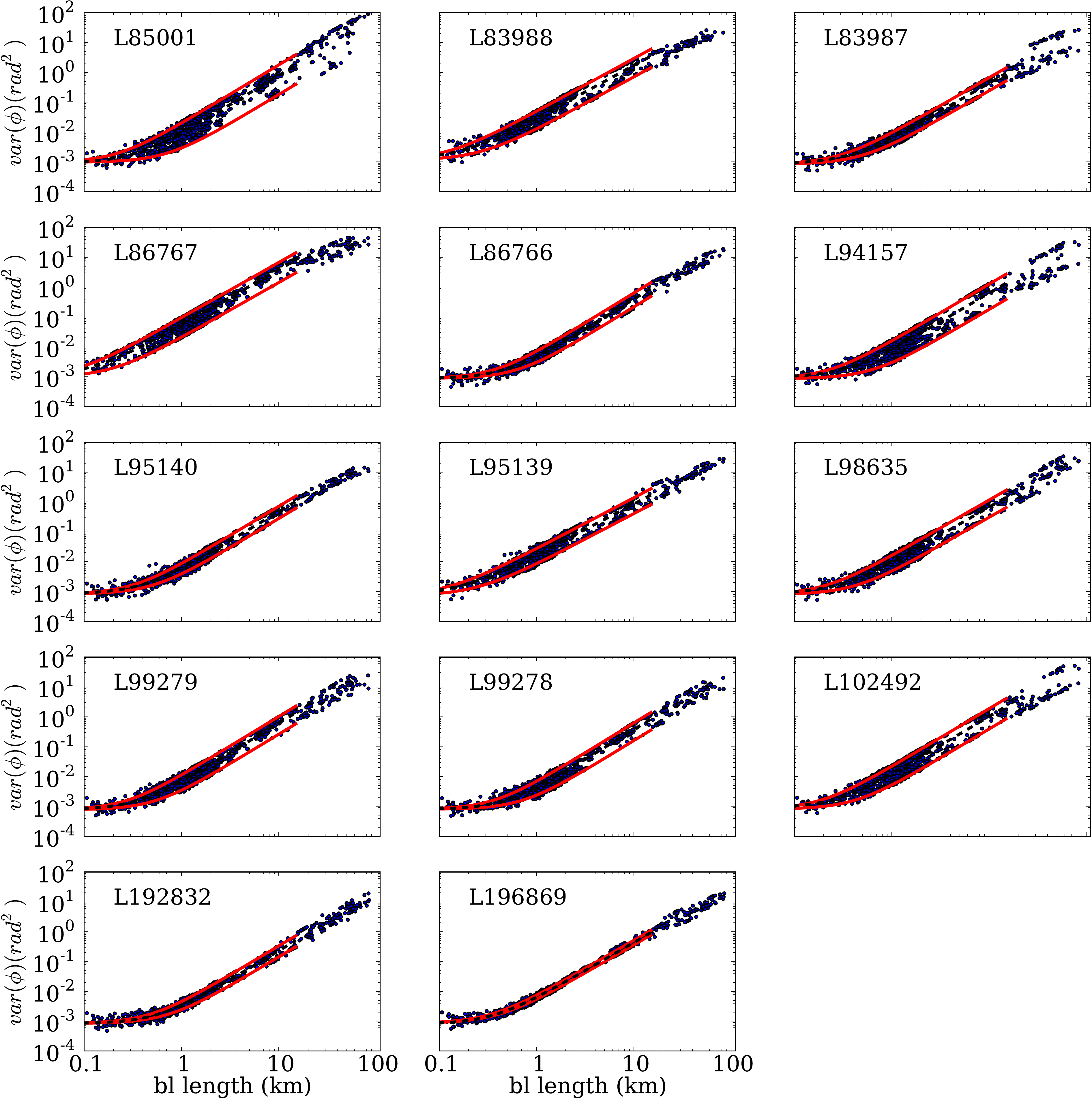}
\caption{Structure functions of all observations. The lines are the major and minor
  axis projections of the fitted function \eqref{eq:2DKolmogorov}.}
\label{fig:structure_all2}
\end{figure*}

\section{Results}
\subsection{Structure Functions Fit}
\begin{figure}[h]
\noindent\includegraphics[width=\hsize]{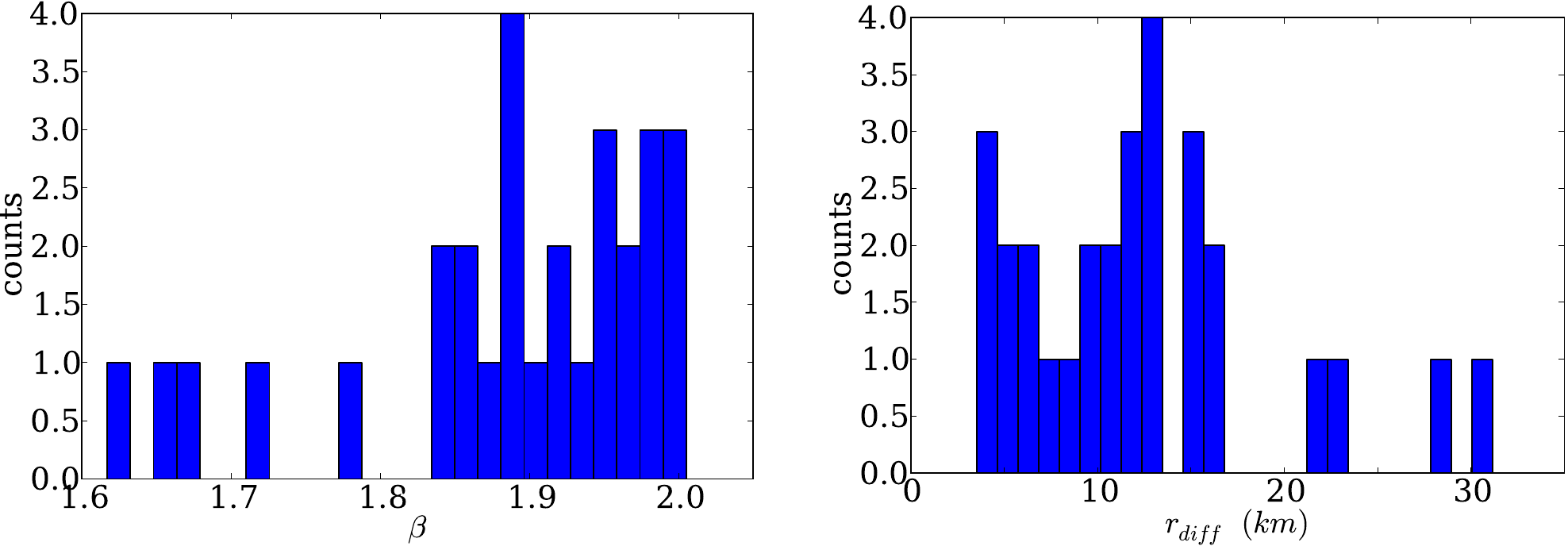}
\caption{Left: histogram of measured values for $\beta$. Right: histogram of
  measured values for diffractive scale $r_{diff}$. }
\label{fig:beta_histo}
\end{figure}

Figures \ref{fig:structure_all1} and \ref{fig:structure_all2} show the phase
structure functions of all 29 observations with the  results of the fits superimposed.
We summarize our results of the fits described in section \ref{section:phasestructure}
in table \ref{tab:fitresults}. Figure \ref{fig:beta_histo} a,b show 
histograms of the main characteristics of the ionosphere, namely the fitted
slope $\beta$ and $r_{diff}$. We notice that on
average the value for $\beta$ is larger than the pure Kolmogorov value of
$5/3$. The average value for $\beta$ is 1.89 with a standard deviation of 0.1.  A priori there is no reason why the power index of the ionospheric structures should be exactly $5/3$ , which is the
value derived for turbulence in the lower atmosphere \citep{rufenach}. The
higher index is likely to be due to non-turbulent structures (for
example Traveling Ionospheric Disturbances (TIDs)\citep{velthoven} or density
ducts \citep{mwa_field_lines}) in the ionosphere. The contribution of the
wavelike TIDs to the structure function is a power law with power 2.0, which
from figure \ref{fig:beta_histo} appears to be the cut-off value for
$\beta$. Besides, both the turnover at long baselines that is observed in some
observations and the orientation dependent band-like structure are also
consistent with being caused by large-scale, coherent fluctuations such as medium scale TIDs.

The standard deviation measured from the noise floor is fairly constant per observation and
on average $0.9\,$mTEC. This is the minimal level of accuracy on differential
TEC that we can achieve on data of a bright calibrator. Improving the model of the
instrument and the sky can further increase the accuracy. The distribution of the diffractive scale $r_{diff}$ values varies between 3
and 30 km. 

\subsection{Diffractive scale}\label{sec:image_noise}

\begin{figure}
\noindent\includegraphics[width=\hsize]{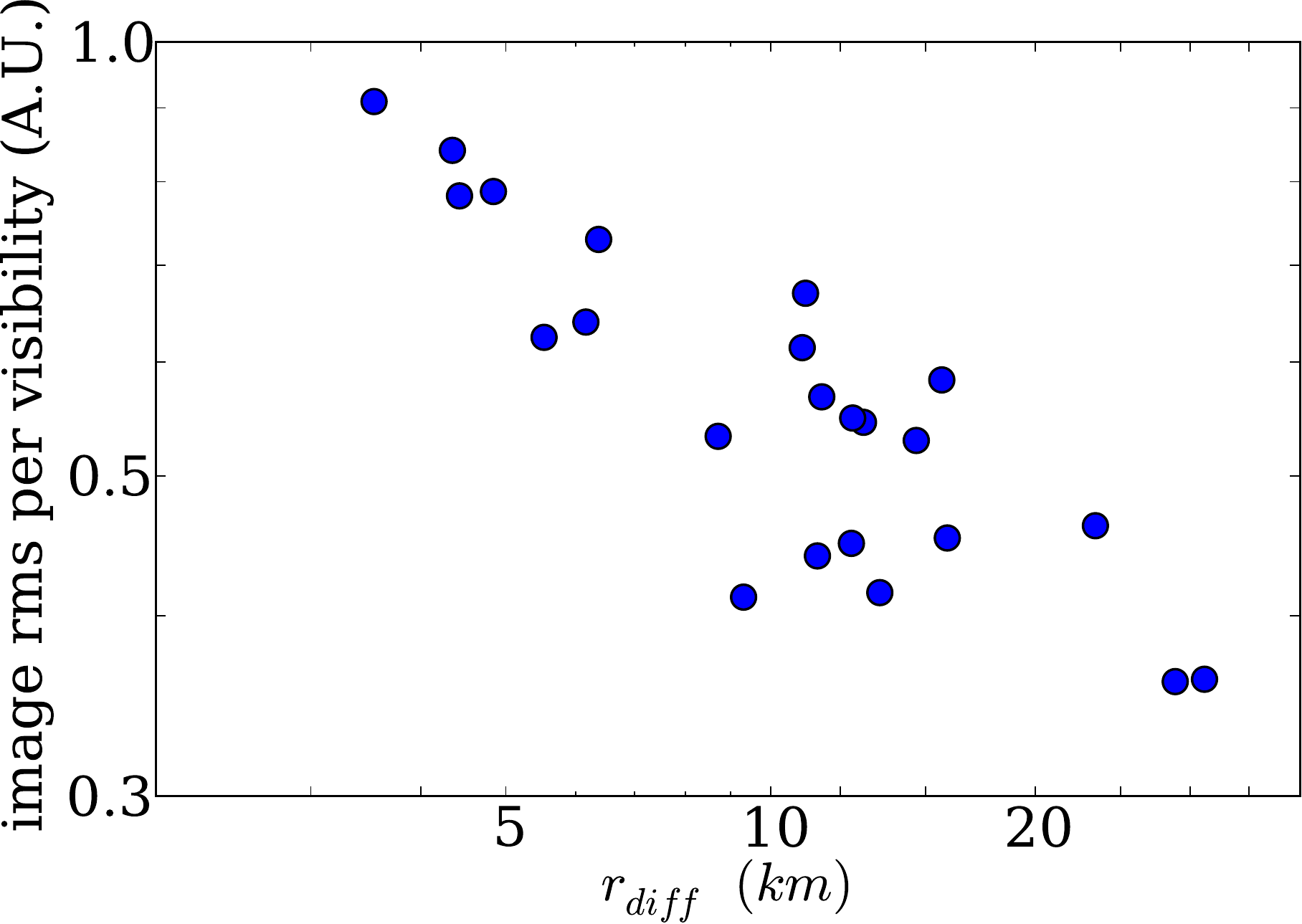}
\caption{Image rms versus measured diffractive scale. The image rms was scaled with
  $\sqrt{\#\,visibilities}$ and calculated after one round of direction
  independent self calibration using a four component model of the central
  source only. }
\label{fig:rmsvss0}
\end{figure}

A small diffractive scale corresponds to large phase
fluctuations over the field of view. The time coherence of the signal is
shorter for smaller diffractive scales, depending also on the
baseline length \citep{vedantham}. To compensate for these fluctuations, it is
needed to calibrate the phases in many directions on short timescales. The
number of directions and the time resolution of the solutions is limited by
the available source flux and number of independent data points \citep{bregman2012}. The $r_{diff}$ appears to be a good measure of the
ionospheric quality for radio interferometric imaging of the night. For EoR
purposes  it was shown in \citet{vedantham} that a diffractive scale larger
than 5 km at 150 MHz is sufficient. This corresponds to $90 \%$ of the observations. In extreme cases,
when the diffractive scale becomes smaller than the Fresnel scale (about 300m
at 150 MHz), we get diffractive amplitude and phase scintillations. Such
conditions have been observed with LOFAR, but not for the observations under consideration. For the observations under consideration we
noticed a large night to night variation in the rms image noise of data only calibrated with direction-independent calibration. We measured the image
noise as the root mean square of the pixel values in source free
regions at the edges of the images, where each image was constructed from
the calibrated visibilities of a single subband.  Interpolation of a polynomial
fit on the rms per frequency resulted in a rms value at the lower edge of the
frequency range, where the ionospheric effects are most severe. For the images
the full 8 hour dataset, if available, was used. To compare observations with
different lengths, the measured image noise values were scaled with the square
root of the number of visibilities. This scaling is only approximately
correct, since the noise in the images is far from thermal. However,  the
number of visibilities between different observations differ only by $30\%$ at
most and we only qualitatively investigate the correlation between diffractive
scale and image noise, justifying using this approximation here. In order to
test if the night to night rms variation could be explained by variation in
the ionospheric conditions, we plotted in figure \ref{fig:rmsvss0} the
rescaled image rms versus the measured diffractive scale. The rms values are
rescaled to an arbitrary scale between 0 and 1. Although the
scatter is large, there is clear evidence of an inverse correlation between
the diffractive scales and image noise. A quantitative analysis and comparison to
\citet{vedantham}, who computed the theoretical residual ionospheric noise
after self calibration, is left for future work. Figure \ref{fig:rmsvss0} is
merely an illustration of the anti-correlation between image noise and
diffractive scale, therefore, the rms values are
rescaled to an arbitrary scale between 0 and 1.

\subsection{Anisotropy}\label{sec:anisotropy}

\begin{figure}
\noindent\includegraphics[width=\hsize]{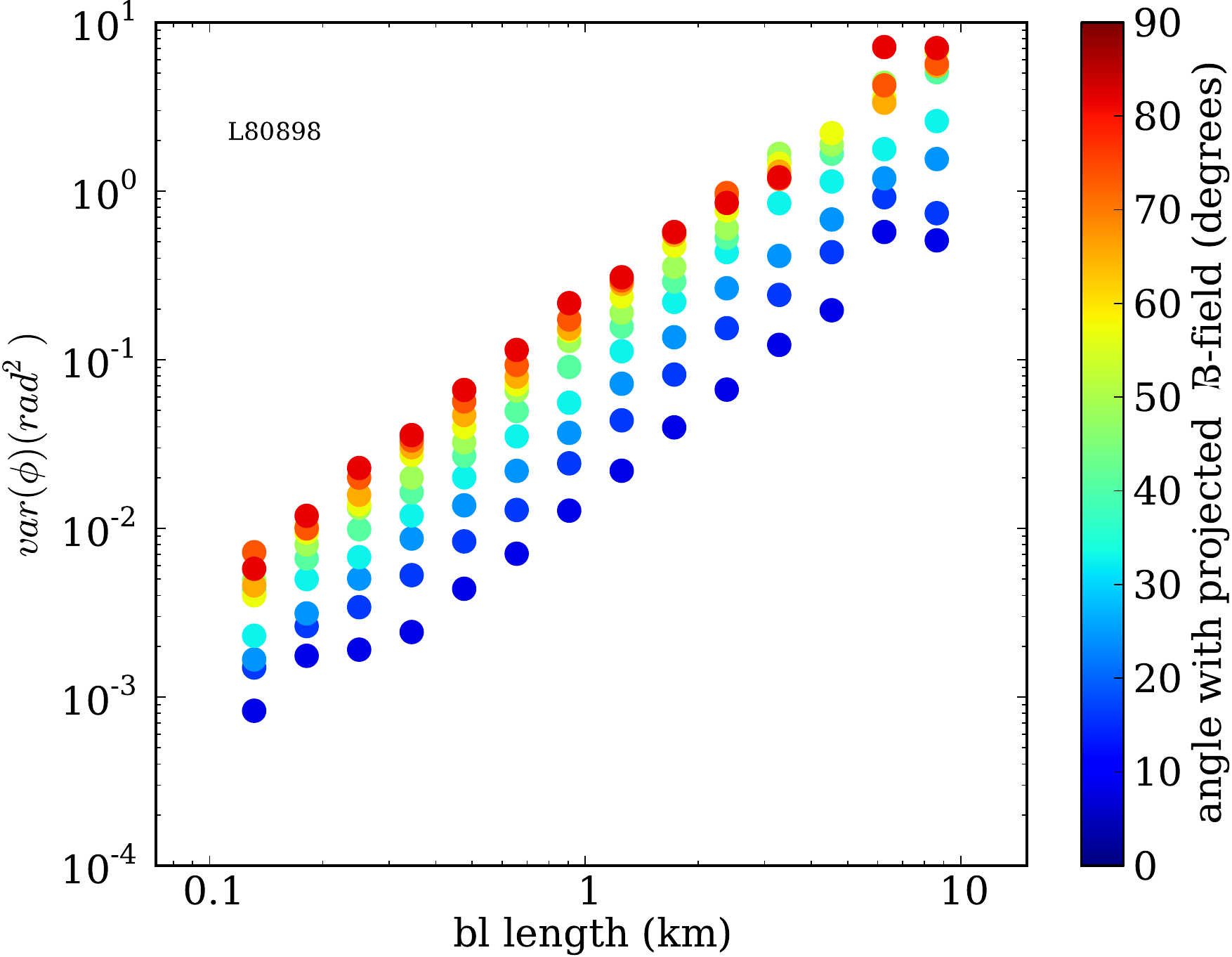}
\caption{Phase structure function of a single observation, with the data binned
  according to the angle with respect to the projected Earth
  magnetic field, given in radians with the color bar. The diffractive scale is evidently
  smaller, thus the points are higher,  for perpendicular baselines in comparison to the parallel.}
\label{fig:field_aligned_structure}
\end{figure}

We investigated the level of anisotropy in the ionospheric structure as well as
the angle of the major axis with respect to the North meridian. The level of
anisotropy, i.e ratio of the major and the minor axis, in table \ref{tab:fitresults} ranges from 1 to
4. Earlier
observations \citep{spencer,singleton,wheelon} suggest that the ionospheric irregularities are aligned and elongated
along the Earth magnetic field lines, which would result in a larger
diffractive scale for the field aligned baselines. Taking into account the
viewing direction in the direction of 3C196, one gets a perspective view of the
magnetic field lines when projected on a given ionospheric altitude. This
method of projecting the field lines is discussed in \citet{mwa_field_lines}.
Since the angle of the
baseline with respect to the projected magnetic field lines changes with time, a single
variance per baseline cannot be calculated anymore. Instead, after removing
a global trend per baseline, the TEC data were binned in 2 dimensional angle
and length bins. We removed the global trend explicitely, where before it was done
implicitly by making use of the phase variance, since different
baselines contribute to the same bin. We used the World Magnetic Model for modeling the magnetic
field lines \citep{wmm}. An example of the variance per bin for a single
observation, in which the observed field aligned anisotropy was high,
is shown in figure
\ref{fig:field_aligned_structure}. For many, but not all, observations that do
show anisotropy, the
ionospheric irregularities seem to be indeed elongated along the magnetic field
lines, with relative differences in diffractive scale in the field aligned and
perpendicular direction up to a factor 6. In the last column of table
\ref{tab:fitresults} we give the ratio of the field aligned and perpendicular
diffractive scales. For about $60\%$ of the observations this ratio is larger
than 2, indicating some field aligned irregularities. This
correlation with the magnetic field could well be related to the \emph{density
  ducts} as observed by \citet{mwa_field_lines}. Their observations show
elongated field aligned structures that move  with low speeds.  If such structures were present during
the observations under consideration, they would indeed lead to similar
structure functions.

\begin{figure}

\noindent\includegraphics[width=\hsize]{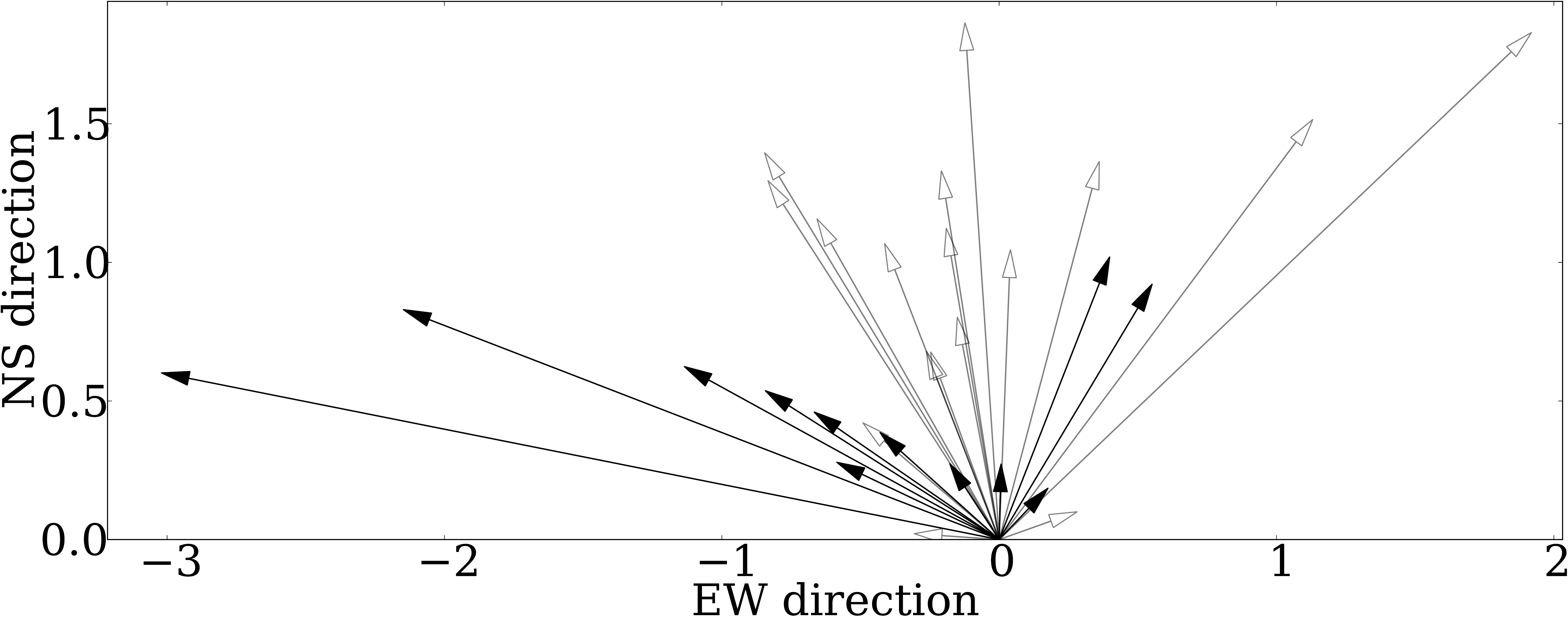}
\caption{Direction and size of anisotropy for all observations. Closed arrows:
  observations that do not show
  Earth magnetic field aligned structures. Open arrows: Observations that have a
  field aligned structure ratio of at least 2. The length of the vectors is defined as
  $r_{diff}^{maj}/r_{diff}^{min}-1$.}
\label{fig:anisotropy_angle_distribution}
\end{figure}

Another contributor to the observed anisotropy could be medium size Traveling
Ionospheric Disturbances (mTIDs)\citep{velthoven}. TIDs are in general not
field aligned, but they could well be the explanation for those observations
that do show anisotropy, but for which the anisotropy is not field aligned.  Baselines that are parallel to the wavefront
of such waves are insensitive to their contribution to the
phase structure. The structure function of a single wave has a slope of
2.0 until the  turnover point around half a wavelength. Since no significant
turnover is observed below $80$ km, the wavelengths of these waves have to be larger than $150$ km. The observed
anisotropies would correspond to amplitudes of the waves of the order of
$0.1-0.5$ TECU, depending on the level of anisotropy and the assumed wavelength. These values are consistent with typical
amplitudes and wavelengths of mTIDs \citep{velthoven}. This is also in
agreement with the distribution of values for $\beta$, which appear to cut-off
at a value of 2.0. In this case, the angle of the major axis
of the 2 dimensional structure function gives an indication of the traveling
direction of such waves. The angles of the major axes with respect to the
North meridian  are plotted in
figure \ref{fig:anisotropy_angle_distribution}. The closed arrows correspond to the observations where no correlation with the
magnetic field was observed,  i.e. where the ratio of the field aligned and
perpendicular diffractive scales was smaller than 2. The open arrows denote
the observations where 
field aligned structures were present. It should be stressed, however, that the vectors are calculated
using the original 2D structure functions, so without using the Earth magnetic
field projection. In this plot the length of the
vectors is determined by the level of anisotropy minus 1: $ r_{diff}^{maj}/r_{diff}^{min}-1$,
such that a zero length vector corresponds to no anisotropy.
We do not observe a preferred direction of the anisotropy of the non field
aligned structures. 

The fact that the ionospheric
structures are anisotropic with the irregularities in many cases geomagnetically aligned,
probably has minor impact on the original purpose of our research, namely investigating calibration
strategies and estimating noise characteristics. However, the two different
diffractive scales need be taken into account in the calculations of estimated
noise. Also, a model of the ionosphere for calibration purposes, e.g. with a
method like SPAM \citep{intema}, could benefit from the knowledge that the
structures are field aligned structures and thus tilted with respect to the Earth's surface.


\section{Conclusion}

We have shown that LOFAR is able to measure with
an accuracy better than $1\,$mTEC the differential TEC values on
spatial scales between 1 and 100 km,  when using the calibration phases with a time resolution of 10 s in the
direction of a bright calibrator. Using a better sky and instrument model
during calibration could improve this accuracy. 

We measured the spatial phase structure above LOFAR of many nights in the winters
of 2012/2013 and 2013/2014. We observed a power law behavior over a long range of baseline lengths, between 1 and 80 km. The
average slope is $1.89$ with a one standard deviation spread of $0.1$, in
general larger than the $5/3$ expected for Kolmogorov turbulence. In some
observations, we have tentatively detected a turnover in the phase structure
function at baseline lengths larger than 80 km. Both the turnover and the large
$\beta$ values are an indication of the contribution of coherent fluctuations,
like for example traveling waves, to the structure function.

There is a large night-to-night variation of the diffractive scale,
which directly translates into image noise variation, if the directional
dependence of the ionospheric distortion is not properly accounted for in the
calibration. Therefore, the diffractive scale could serve as a measure of the
ionospheric quality of an observation. Theoretical predictions of the expected image
noise dependence on ionospheric diffractive scales are discussed in
\citet{vedantham}. When minimal
residual noise is required, such as for the LOFAR EoR project, it is important
to assign an ionospheric quality in order to decide which observations are useful.  The measured 
diffractive scales at $150\,$MHz range between $3$ and
$30\,$km. \citet{vedantham} have argued that for the LOFAR EoR measurement a
minimal $r_{diff}$ of $5$ km is needed. This is the case for $90\%$ of the observations discussed here.
Furthermore, for the generation of a sky model used in subsequent direction
dependent calibration, only the best nights, with a $r_{diff}>20 \,$km, are
used, such that image distortion at the highest resolution is minimal. 

We observed an anisotropy ratio of the ionospheric structure ranging between
1 (no anisotropy) and 6, that is in many, but not all, cases aligned along
the  Earth magnetic field. The field aligned anisotropy is reminiscent of
the observations of large field aligned structures in
\citet{mwa_field_lines}. If not Earth magnetic field aligned, the observed
anisotropy could again also be explained by a contribution from traveling
waves, where the shortest diffractive scales
are measured along the wave propagation direction.

It should be noted that the results discussed here only consider nighttime
data. Ionospheric effects during daytime and sunrise and sunset are expected
to be different due to the solar irradiation. Also, all observations were
done in wintertime. It is likely that seasonal variations have an effect on the ionospheric structure function.

\begin{acknowledgments}
This research was supported by grants from the Netherlands Organization for
Scientific Research (TOP grant no 614.001.005) and from the European Research
Council (grant no 339743 (LOFARCORE)). 

All data presented here are made available upon request by the corresponding author.
\end{acknowledgments}

%

\afterpage{\clearpage\thispagestyle{empty}
\begin{sidewaystable*}[p]
  \caption{Parameters of all observations. In the last column (\emph{fa}) the
    ratio of field aligned and field perpendicular $r_{diff}$ is given.}
  \label{tab:fitresults}
  \begin{tabular}{ccccS[table-format=1.2,separate-uncertainty,table-figures-uncertainty=1]
    S[table-format=2.1,separate-uncertainty,table-figures-uncertainty=1]
S[table-format=2.1,separate-uncertainty,table-figures-uncertainty=1]
S[table-format=2.1]S[table-format=1.2]S[table-format=3.1,separate-uncertainty,table-figures-uncertainty=1]c}
    \hline
    {Id} & {Date} & {Start time} & {Length} & {$\beta$} & {Major} & {Minor}&
    {$r_{diff}$}& {$\sigma$}  &{$\alpha$} & {fa}\\
    & & {(UTC)} & {(hr)} &  & {(km)} & {(km)}&
    {$(km)$}& {$(mTEC)$}  & &\\
    \hline
L78444 & 2012/11/30 & 23:04:41 & 8 & 1.92 \pm 0.04 &  12.8 \pm 0.2 &  10.4 \pm 0.7 &  11.4 & 1.00 & 147.0 \pm 8.6 & 1.8\\
L79324 & 2012/12/06 & 22:41:05 & 8 & 1.88 \pm 0.02 &   5.8 \pm 0.2 &   3.6 \pm
0.1 &   4.3 & 0.91 & 158.9 \pm 1.9 & 3.2 \\
L79344 & 2012/12/09 & 22:29:17 & 8 & 1.96 \pm 0.02 &  13.8 \pm 0.6 &   6.3 \pm 0.2 &   8.7 & 1.05 & 118.8 \pm 0.8 & 1.2\\
L80273 & 2012/12/12 & 22:17:30 & 8 & 1.98 \pm 0.01 &  22.8 \pm 0.5 &   9.3 \pm 0.3 &  13.3 & 0.97 & 147.2 \pm 1.1 & 2.3\\
L80508 & 2012/12/16 & 22:01:46 & 8 & 2.00 \pm 0.01 &  23.1 \pm 0.7 &   5.8 \pm 0.2 &   9.3 & 1.04 & 101.3 \pm 0.4 & 0.8\\
L80898 & 2012/12/19 & 21:49:58 & 8 & 1.67 \pm 0.02 &   5.5 \pm 0.2 &   2.8 \pm 0.1 &   3.5 & 0.46 & 2.2 \pm 1.4 & 5.8\\
L80897 & 2012/12/21 & 21:42:07 & 8 & 1.86 \pm 0.03 &  25.3 \pm 0.1 &  21.8 \pm 1.5 &  23.4 & 1.41 & 43.6 \pm 17.3 & 1.6\\
L80895 & 2012/12/28 & 21:14:35 & 8 & 1.62 \pm 0.04 &  15.0 \pm 0.1 &   4.2 \pm 1.2 &   6.4 & 0.61 & 46.4 \pm 0.8 &4.3 \\
L82609 & 2012/12/29 & 21:10:39 & 8 & 1.89 \pm 0.01 &  33.7 \pm 0.8 &  28.8 \pm 0.6 &  31.1 & 1.01 & 1.5 \pm 5.5 & 0.9\\
L82655 & 2013/01/01 & 20:58:51 & 8 & 1.98 \pm 0.01 &  18.0 \pm 0.4 &   9.5 \pm 0.3 &  12.4 & 1.13 & 122.5 \pm 1.9 & 1.5\\
L80893 & 2013/01/04 & 20:47:04 & 8 & 1.84 \pm 0.04 &   6.1 \pm 0.0 &   5.1 \pm 0.2 &   5.5 & 1.24 & 70.4 \pm 8.0 & 2.9\\
L80892 & 2013/01/09 & 20:27:24 & 8 & 1.94 \pm 0.01 &  35.4 \pm 1.1 &  23.9 \pm 0.5 &  28.8 & 1.01 & 131.9 \pm 3.5 & 1.1\\
L80891 & 2013/01/11 & 20:19:32 & 8 & 1.91 \pm 0.02 &   7.8 \pm 0.2 &   5.1 \pm 0.1 &   6.2 & 1.23 & 115.5 \pm 3.4 & 1.3\\
L83991 & 2013/01/18 & 19:52:01 & 8 & 1.84 \pm 0.04 &   8.7 \pm 0.5 &   3.1 \pm 0.1 &   4.4 & 0.61 & 36.7 \pm 0.7 & 2.7\\
L84999 & 2013/01/21 & 19:40:13 & 8 & 1.66 \pm 0.03 &  19.2 \pm 0.6 &   7.6 \pm 0.4 &  10.9 & 0.72 & 148.8 \pm 1.2 & 3.2\\
L85001 & 2013/01/26 & 19:20:33 & 8 & 1.97 \pm 0.01 &  24.0 \pm 0.6 &   7.5 \pm 0.2 &  11.3 & 0.97 & 111.1 \pm 1.3 & 1.3\\
L83988 & 2013/01/30 & 21:04:50 & 6 & 1.78 \pm 0.03 &  12.2 \pm 0.4 &   5.5 \pm 0.4 &   7.4 & 1.12 & 150.4 \pm 3.1 & 2.4\\
L83987 & 2013/02/01 & 18:56:58 & 8 & 1.88 \pm 0.02 &  21.5 \pm 0.5 &  12.5 \pm 0.3 &  15.7 & 0.83 & 169.4 \pm 2.5 &4.2 \\
L86767 & 2013/02/07 & 18:33:23 & 8 & 1.85 \pm 0.02 &   8.3 \pm 0.3 &   3.6 \pm 0.1 &   4.8 & 0.95 & 14.8 \pm 1.1 & 3.9\\
L86766 & 2013/02/08 & 18:29:27 & 8 & 2.00 \pm 0.01 &  21.4 \pm 0.4 &  12.5 \pm 0.7 &  15.9 & 0.88 & 124.6 \pm 4.0 & 1.6\\
L94157 & 2013/02/15 & 18:01:56 & 8 & 1.92 \pm 0.01 &  24.6 \pm 1.3 &   8.9 \pm 0.2 &  12.8 & 0.86 & 176.2 \pm 1.0 & 4.6\\
L95140 & 2013/02/22 & 17:34:24 & 8 & 1.96 \pm 0.01 &  19.2 \pm 1.0 &  11.9 \pm 0.3 &  14.6 & 0.83 & 160.0 \pm 4.0 & 2.1\\
L95139 & 2013/02/23 & 17:30:28 & 8 & 1.71 \pm 0.03 &  16.9 \pm 0.4 &   8.3 \pm
0.4 &  11.0 & 0.71 & 158.9 \pm 2.0 & 3.2\\
L98635 & 2013/03/03 & 16:59:01 & 8 & 1.88 \pm 0.02 &  18.9 \pm 0.6 &   9.2 \pm 0.1 &  12.4 & 0.79 & 170.4 \pm 1.0 & 2.9\\
L99279 & 2013/03/07 & 17:43:17 & 6 & 1.99 \pm 0.01 &  19.7 \pm 0.4 &   9.9 \pm 0.2 &  12.8 & 0.77 & 21.4 \pm 1.0 & 1.3\\
L99278 & 2013/03/08 & 17:39:21 & 6 & 2.01 \pm 0.01 &  25.0 \pm 1.1 &  12.6 \pm 0.3 &  16.3 & 0.80 & 30.9 \pm 0.6 & 0.9\\
L102492 & 2013/03/12 & 17:23:38 & 6 & 1.95 \pm 0.02 &  16.3 \pm 0.6 &   7.2 \pm 0.4 &   9.9 & 0.81 & 171.1 \pm 4.6 & 2.2\\
L192832 & 2013/12/15 & 23:06:38 & 6 & 1.95 \pm 0.01 &  27.2 \pm 0.4 &  17.6 \pm 1.2 &  21.3 & 0.84 & 130.6 \pm 2.4 & 2.2\\
L196869 & 2014/01/04 & 21:48:00 & 6 & 1.89 \pm 0.02 &  17.2 \pm 0.5 &  14.2 \pm 0.3 &  15.6 & 0.86 & 93.9 \pm 6.5 & 2.1\\
\hline
  \end{tabular}
\end{sidewaystable*}
}

\end{article}
%
%
%
%
%
%
%
%


\end{document}